\documentclass[journal]{new-aiaa}
\usepackage[utf8]{inputenc}
\usepackage{textcomp}

\usepackage{graphicx}
\usepackage{amsmath}
\usepackage[version=4]{mhchem}
\usepackage{siunitx}
\usepackage{longtable,tabularx}
\usepackage{times}
\usepackage[numbers]{natbib}
\usepackage{acronym}
\usepackage{fancyhdr}
\usepackage{verbatim}
\usepackage[numbers]{natbib}
\usepackage{bm}
\usepackage{commath}
\usepackage{booktabs}
\usepackage{enumitem}
\usepackage{algorithm}
\usepackage[indLines=true]{algpseudocodex}

\DeclareMathOperator*{\argmax}{arg\,max}
\DeclareMathOperator*{\infimum}{inf}

\DeclareSIUnit[number-unit-product = ]\percent{\char`\%}
\DeclareSIUnit\lu{LU}
\DeclareSIUnit\tu{TU}
\DeclareSIUnit\arcsec{arcsec}

\acrodef{ssa}[SSA]{space situational awareness}
\acrodefindefinite{ssa}{an}{a}
\acrodef{engmf}[EnGMF]{ensemble Gaussian mixture filter}
\acrodefindefinite{engmf}{an}{an}
\acrodef{glmbf}[GLMBF]{generalized labeled multi-Bernoulli filter}
\acrodef{glmb}[GLMB]{generalized labeled multi-Bernoulli}
\acrodef{lmb}[LMB]{labeled multi-Bernoulli}
\acrodefindefinite{lmb}{an}{a}
\acrodef{dro}[DRO]{distant retrograde orbit}
\acrodef{nrho}[NRHO]{near-rectilinear halo orbit}
\acrodefindefinite{nrho}{an}{a}
\acrodef{lto}[LTO]{lunar transfer orbit}
\acrodefindefinite{lto}{an}{a}
\acrodef{llo}[LLO]{low lunar orbit}
\acrodefindefinite{llo}{an}{a}
\acrodef{pdf}[PDF]{probability density function}
\acrodef{so}[SO]{space object}
\acrodef{aso}[ASO]{anthropogenic space object}
\acrodef{eci}[ECI]{Earth-centered inertial}
\acrodefindefinite{eci}{an}{an}
\acrodef{mci}[MCI]{Moon-centered inertial}
\acrodefindefinite{mci}{an}{a}
\acrodef{srukf}[SR-UKF]{square root unscented Kalman filter}
\acrodef{ospa}[OSPA]{optimal subpattern assignment}
\acrodefindefinite{ospa}{an}{an}
\acrodef{srp}[SRP]{solar radiation pressure}
\acrodefindefinite{srp}{an}{a}
\acrodef{kde}[KDE]{kernel density estimation}
\acrodef{gmm}[GMM]{Gaussian mixture model}
\acrodef{rfs}[RFS]{random finite set}
\acrodefindefinite{rfs}{an}{a}
\acrodef{mtt}[MTT]{multi-target tracking}
\acrodefindefinite{mtt}{an}{a}
\acrodef{crtbp}[CRTBP]{circular restricted three-body problem}
\acrodef{rmse}[RMSE]{root-mean-square error}
\acrodef{qoi}[QoI]{quantity of interest}

\title{Multi-Fidelity Uncertainty Propagation with Model Adaptation to Local Cislunar Dynamics}

\author{Cedric Petion\footnote{Graduate Research Assistant, Department of Aerospace Engineering and Engineering Mechanics; cpetion@utexas.edu (Corresponding Author).}, Benjamin L.\ Reifler\footnote{Postdoctoral Fellow, Oden Institute for Computational Engineering \& Sciences. Member AIAA.}, and Brandon A.\ Jones\footnote{Associate Professor, Department of Aerospace Engineering and Engineering Mechanics. Associate Fellow AIAA.}}
\affil{The University of Texas at Austin, Austin, TX 78712}

\begin{document}

\maketitle


\section{Introduction} \label{sec:intro}

\lettrine{T}{he} cislunar regime is increasingly receiving attention from governmental and commercial entities, with the growing number of planned missions expected to increase the number of \acp{aso} in the region. This proliferation of \acp{aso}, along with cislunar space's size and illumination conditions that contribute to sparse measurements and the chaotic multi-body dynamics that complicate uncertainty propagation, presents a need for advances in cislunar \ac{ssa}~\cite{holzinger_primer, wishnek_holzinger_handley, bolden_et_al, bhadauria2022, petion_reifler_ECSD, cedric_amos}. This paper presents an adaptive multi-fidelity uncertainty propagation method that dynamically adjusts the included perturbing forces based on position in cislunar space, minimizing computation time while maintaining a prescribed modeling accuracy. The proposed adaptive method is then integrated into \iac{mtt} framework to reduce the computational cost of track prediction without sacrificing accuracy, which is important for managing the growing number of objects in cislunar space.

An essential capability for cislunar \ac{ssa} is the ability to track a large number of \acp{aso}, which may include both operational spacecraft and space debris. \Acl{mtt} algorithms provide a framework to achieve this, but incur a computational cost for the prediction step proportional to the cost of propagating a single object's state \ac{pdf} times the number of hypothesized tracks. Thus, to maintain computational tractability, efficient orbit uncertainty propagation methods must be employed. The adaptive multi-fidelity orbit propagation method developed in this paper is applied to an \ac{engmf}, a particle-based filter that is robust to sparse measurements~\cite{Anderson1999, Yun2022, reifler_popov_jones_zanetti_2023}. The \ac{engmf} is used in a \ac{glmbf}, which is a multiple hypothesis multi-target filter based on labeled \acp{rfs}~\cite{Vo2013, Vo2014, Vo2017}.

Multi-fidelity methods were first proposed for orbit uncertainty propagation by~\cite{jones_weisman}. This approach propagates a particle-based representation of a state-space \ac{pdf} using a fast-to-compute but possibly inaccurate low-fidelity model, and then identifies a small subset of “important” samples to re-propagate
with an accurate yet computationally expensive high-fidelity model. These low- and high-fidelity particles are then used in a stochastic collocation procedure to produce an approximate high-fidelity solution (designated the multi-fidelity solution) for each point in the full particle ensemble. The multi-fidelity solution then has accuracy close to that of the high-fidelity model but runtime close to the low-fidelity model.

The accuracy and runtime of the multi-fidelity solution are contingent on the low- and high-fidelity models that are selected. For cislunar applications, the optimal model pair varies throughout cislunar space as the relative contributions of Earth gravity, Lunar gravity, and perturbing forces change throughout different domains. Thus, when propagating a \ac{pdf} through cislunar space with the multi-fidelity method, the optimal model pair may change over time and space and may not be known a priori.

Our solution is an adaptive approach to model selection that varies the perturbations included in the high-fidelity model as a function of position in cislunar space. We pose the issue as an optimization problem, with the objective being to minimize runtime subject to an upper bound on the expected error in acceleration magnitude. The solution is obtained using a precomputed library of recommended gravity field terms from \cite{McArdle} and \cite{McArdle_GNC}. This work was previously presented at the 9th European Conference on Space Debris as \cite{petion_reifler_ECSD} and advances prior research from~\cite{ben_amos} that used a limited set of three models: one pair of models near the Earth, one near the Moon, and one in all other regimes, as well as the work of~\cite{trevor_cislunar_mf} that examined non-adaptive cislunar multi-fidelity propagation.

Background on the multi-fidelity method, the \ac{engmf}, and the \ac{glmbf} are provided in Sections \ref{sec:background_mf}, \ref{sec:background_engmf}, and \ref{sec:background_glmbf}, respectively. The adaptive multi-fidelity model selection method is outlined in Section \ref{sec:methods}. Lastly, numerical results for simulated test cases are presented in Section \ref{sec:numeric_simulations}, which assesses the propagation accuracy, tracking accuracy, and runtime for cislunar \ac{ssa} scenarios.
\section{Background} \label{sec:background}

\subsection{Multi-Fidelity Orbit Uncertainty Propagation} \label{sec:background_mf}
The multi-fidelity method employed in this paper was first proposed in \cite{mf_1} and \cite{mf_2} and later extended to orbit uncertainty propagation by \cite{jones_weisman}. The approach used in this paper is bi-fidelity, with a low-fidelity model that is intended to be fast to evaluate but may possess relatively poor accuracy, and a high-fidelity model with greater accuracy at the expense of increased computation time.

Reference~\cite{jones_weisman} suggests using a general perturbations propagator for the low-fidelity model to provide analytic or semi-analytic solutions that are rapid to compute. However, in cislunar space, where the gravities of the Earth and the Moon are of similar magnitudes, general perturbations methods, which typically assume small third-body perturbations (if any at all), often fail to provide sufficiently accurate solutions. Therefore, in this work, the low-fidelity model is a special perturbations propagator that numerically integrates the differential equations for a limited, and thus relatively fast to evaluate, force model. The high-fidelity model is also a special perturbations propagator, but with a force model containing substantially more perturbations that are relatively slow to compute. See Section \ref{sec:methods} for an in-depth discussion on the model selection.

Conceptually, the multi-fidelity method works by using a low-fidelity dynamics model to propagate a set of particles (such as Monte Carlo samples or sigma points), and then re-propagates a limited subset of particles that are dynamically important with a high-fidelity dynamics model and uses these “important” samples to adjust the low-fidelity samples via stochastic collocation.

The remainder of this section provides an abbreviated explanation of the multi-fidelity method in an orbit propagation context. For a complete derivation and discussion, see \cite{mf_1}, \cite{mf_2}, and \cite{jones_weisman}. Let $(\Omega,\Sigma,\mathcal{P})$ be a probability space with the sample space $\Omega$, $\sigma$-algebra $\Sigma$, and probability measure $\mathcal{P}$. For a stochastic system with $d$ independent random inputs, denote the vector of random inputs by $\bm{\xi}: \Omega \to \Gamma^d \subseteq \mathbb{R}^d$ defined on $(\Omega,\Sigma,\mathcal{P})$. The support $\Gamma^d$ is determined by the probability distribution of $\bm{\xi}$. For $d$-dimensional Gaussian random variables, $\Gamma^d = \mathbb{R}^d$, whereas bounded distributions yield compact support. In this work, a fixed realization of $\bm{\xi}$ corresponds to a Monte Carlo sample from the probability distribution of initial conditions and may also be concatenated with a realization of the process noise acceleration. We seek a solution to the orbit uncertainty propagation problem governed by the dynamical system
\begin{equation}\label{eq:dyn_sys}
\begin{cases}
\dot{\bm{x}}(t,\bm{\xi}) = \bm{f}(t,\bm{x}(t,\bm{\xi}),\bm{\xi}),\\
\bm{x}(t_0,\bm{\xi}) =  \bm{x}_0(\bm{\xi}),
\end{cases}
\quad \mathcal{P}\text{ a.s. in } \Omega,
\end{equation}
where $\bm{f}: [t_0,~t_f] \times \mathbb{R}^{n_x} \times \Gamma^d \to \mathbb{R}^{n_x}$ are the equations of motion, $t \in [t_0,~t_f]$ is the time variable, $\bm{x} \in \mathbb{R}^{n_x}$ is the state variable, and $\bm{x}_0(\bm{\xi})$ is the initial condition. In other words, the uncertainty propagation problem is to quantify the effect of the random inputs $\bm{\xi}$ on the state $\bm{x}(t,\bm{\xi})$.

In the multi-fidelity setting, a low-fidelity model for $\bm{f}$ is denoted by $\bm{f}^L$ and a high-fidelity model by $\bm{f}^H$. We seek a multi-fidelity solution for the \ac{qoi} $\tilde{\bm{x}} \in \mathbb{R}^n$, which in this paper is the orbit state vector, possibly at multiple times, thus
\begin{equation}
    \tilde{\bm{x}}(\bm{\xi}) = 
    \begin{bmatrix}
        \bm{x}(t_1,\bm{\xi})^\intercal &
        \bm{x}(t_2,\bm{\xi})^\intercal &
        \cdots &
        \bm{x}(t_T,\bm{\xi})^\intercal
    \end{bmatrix}^\intercal.
\end{equation}
The reason for defining the \ac{qoi} as the state vector at multiple times will be made apparent shortly. The state vector in this work is the Cartesian position and velocity, i.e., $\bm{x}(t_j,\bm{\xi}) = \left[\bm{r}(t_j,\bm{\xi})^\intercal~~\dot{\bm{r}}(t_j,\bm{\xi})^\intercal\right]^\intercal$, though in general the state may be represented by an orbital element set and $\tilde{\bm{x}}$ may contain other \acp{qoi} that are not state variables.

We use $\bm{f}^L$ and $\bm{f}^H$ to generate low- and high-fidelity \acp{qoi} (propagated samples) $\tilde{\bm{x}}^L(\bm{\xi})$ and $\tilde{\bm{x}}^H(\bm{\xi})$, respectively. Then, for the set of realizations of random inputs $\Xi=\{\bm{\xi}_i\}_{i=1}^m$, define the snapshot matrix of low-fidelity samples
\begin{align} \label{eq:snapshot}
    \bm{X}^L(\Xi) &= 
    \begin{bmatrix}
        \tilde{\bm{x}}^L(\bm{\xi}_1) & \cdots & \tilde{\bm{x}}^L(\bm{\xi}_m)
    \end{bmatrix} \in \mathbb{R}^{n \times m},
\end{align}
and the subspace
\begin{align}
    \mathbb{X}^L(\Xi) &= \operatorname{span}( \bm{X}^L(\Xi) ).
\end{align}
We desire a multi-fidelity approximation provided by a stochastic collocation surrogate
\begin{equation}\label{eq:hf_expansion}
    \tilde{\bm{x}}^H(\bm{\xi}) \approx \hat{\tilde{\bm{x}}}^H(\bm{\xi}) = \sum_{l=1}^r c_l(\bm{\xi})\tilde{\bm{x}}^H(\Bar{\bm{\xi}}_l),
\end{equation}
where $\hat{\cdot}$ denotes an approximate value for the variable $\cdot$, $c_l$ are expansion coefficients, $\Bar{\bm{\xi}}_l$ are the important samples and used as the random inputs for the collocation points in the expansion, and $r$ is the rank of the surrogate with $r \ll m$.

The important samples are identified iteratively, with each iteration's important sample being the point that maximizes the distance to the subspace of previously identified important samples. Let $\Xi^{k-1} = \{\Bar{\bm{\xi}}_1,...,\Bar{\bm{\xi}}_{k-1}\}$ for $k > 1$ be the current set of important samples, initialized with $\Xi^0 = \emptyset$ at $k=1$. The $k$th important sample is
\begin{equation} \label{eq:argmax}
    \Bar{\bm{\xi}}_k = \argmax_{\bm{\xi} \in \Xi} \operatorname{dist}( \tilde{\bm{x}}^L(\bm{\xi}), \mathbb{X}^L(\Xi^{k-1}) ),
\end{equation}
where
\begin{equation} \label{eq:distance}
    \operatorname{dist}( \bm{x}, \mathbb{X} ) \equiv \infimum_{\bm{y} \in \mathbb{X}} \norm{\bm{x} - \bm{y}},
\end{equation}
and
\begin{equation} \label{eq:imp_samp_union}
    \Xi^k = \Xi^{k-1} \cup \{\Bar{\bm{\xi}}_k\}.
\end{equation}
The set of important samples is then $\bar{\Xi} = \Xi^{k} = \{\bar{\xi}_l\}_{l=1}^r$. The maximum number of important samples is limited to the rank of the surrogate, with $r \leq n \ll m$. Orbit state vectors typically have dimension $n = 6$, hence why we augmented $\tilde{\bm{x}}(\bm{\xi})$ with the state trajectory at multiple points in time following the recommendation of \cite{jones_weisman}. Doing so increases the rank of the surrogate, thus permitting surrogates with more important samples which improves the accuracy of the multi-fidelity solution. Consequently, the \ac{qoi} vector is defined by points along a trajectory instead of a single-epoch state.

To solve the discrete optimization problem in Eq.~\ref{eq:argmax}, \cite{mf_1} suggests a greedy approach via the pivoted Cholesky decomposition to produce an approximate solution given by
\begin{equation} \label{eq:cholesky}
    \left[\bm{X}^L\right]^\intercal\bm{X}^L = \bm{A}^\intercal \bm{L}\bm{L}^\intercal\bm{A},
\end{equation}
where $\left[\bm{X}^L\right]^\intercal\bm{X}^L$ is the Gram matrix of the low-fidelity samples with respect to the Euclidean inner product, $\bm{L}$ is lower triangular, and $\bm{A}$ is a pivot matrix.

The expansion coefficients are solved such that the basis of low-fidelity important samples may be used to approximately reconstruct the snapshot matrix. The expansion coefficients $c_l$ must then satisfy
\begin{equation}\label{eq:lf_expansion}
    \tilde{\bm{x}}^L(\bm{\xi}) \approx \hat{\tilde{\bm{x}}}^L(\bm{\xi}) = \sum_{l=1}^r c_l(\bm{\xi})\tilde{\bm{x}}^L(\Bar{\bm{\xi}}_l),
\end{equation}
which parallels Eq.~\ref{eq:hf_expansion}. Reference~\cite{jones_weisman} proposes an iterative approach to determine $r$ by incrementing it until the reconstructed snapshot matrix produced by Eq.~\ref{eq:lf_expansion} is within some user-specified tolerance of the original snapshot matrix from Eq.~\ref{eq:snapshot}.

Expressing Eq.~\ref{eq:lf_expansion} in matrix form and combining it with Eq.~\ref{eq:cholesky}, we obtain
\begin{equation}\label{eq:LL^Tc}
    \bm{L}\bm{L}^\intercal\bm{c} = \bm{X}(\bar{\Xi})^\intercal \bm{X}(\Xi),
\end{equation}
where $\bm{c}$ is a matrix of the coefficients $c_l$ and may be solved using forward / backward substitution. The coefficients can then be used in Eq.~\ref{eq:hf_expansion} to obtain the multi-fidelity solution.

\subsection{The Ensemble Gaussian Mixture Filter} \label{sec:background_engmf}
The \ac{engmf} enables accurate and efficient nonlinear estimation. It parameterizes the estimated state \ac{pdf} as a set of particles and
uses \ac{kde} to avoid particle depletion, allowing it to function with fewer
particles than a particle filter \cite{Anderson1999,Yun2022}.
The presentation of the \ac{engmf} in this section follows that of \cite{ben_amos}.
To initialize the \ac{engmf}, $N$ random samples are drawn from the initial
\ac{pdf}.
The \ac{pdf} is then predicted by propagating each particle forward in time.
To update the \ac{pdf}, we first convert the particles to \iac{gmm} via \ac{kde} using Silverman's rule \cite{silverman1998ch4,Yun2022}:
each particle becomes the mean of \iac{gmm} component with weight $N^{-1}$ and covariance
\begin{equation}
\bm{B}_S=\left(\frac{4}{N\left(d+2\right)}\right)^{\frac{2}{d+4}}\bm{P}\,,\label{e:silverman}
\end{equation}
where $\bm{P}$ is the particles' sample covariance.
The \ac{gmm} is then updated using a filtering algorithm formulated for \acp{gmm}, after which $N$ new particles are sampled from the updated distribution.
The estimated state mean and covariance are likewise computed from the \ac{kde}-based \ac{gmm} representation rather than directly from the particle ensemble.

In this work, the measurement update of the \ac{gmm} is performed using the \ac{srukf} \cite{vandermerwe2001}, which provides accurate nonlinear updates with relatively few particles.
This approach lowers the computational expense associated with the prediction and high-fidelity correction stages, although it increases the per-particle computational cost of the measurement update.
The \ac{engmf} generally performs better for orbit determination when its
particles' states are parameterized by equinoctial orbital elements instead of Cartesian coordinates, because their more linear evolution improves the accuracy of the bandwidth parameter in Eq.~\ref{e:silverman} \cite{broucke1972,Yun2022}.
However, because equinoctial elements are designed to represent two-body orbits,
here we instead use Cartesian coordinates.

\subsection{The Generalized Labeled Multi-Bernoulli Filter} \label{sec:background_glmbf}
The presentation of the \ac{glmbf} in this section follows that of \cite{ben_amos}. The multi-target tracking simulations in Section~\ref{sec:results_tracking} use the joint predict--update formulation of the \ac{glmbf}, which is a closed-form solution to the Bayes multi-target filter recursion based on labeled \acp{rfs} \cite{Vo2013,Vo2014,Vo2017}.
\Iac{rfs} may be thought of as a set of random vectors whose cardinality is also a random variable.
More formally, given a space $\mathcal{X}$, \iac{rfs} is a random variable on $\mathcal{F}\mathopen{}\left(\mathcal{X}\right)\mathclose{}$, the set of finite subsets of $\mathcal{X}$.
A labeled \ac{rfs} is a random variable on $\mathcal{F}\mathopen{}\left(\mathcal{X}\times\mathcal{L}\right)\mathclose{}$, where $\mathcal{L}$ is a discrete label space.
This means that each element in a realization $X$ of a labeled \ac{rfs} is of the form $\left(\bm{x},l\right)$, where $\bm{x}$ is the state and $l$ is the label.
A labeled \ac{rfs} realization may not contain duplicate labels, and this is enforced using the distinct label indicator $\Delta\mathopen{}\left(X\right)\mathclose{}=\delta_{\left|X\right|}\mathopen{}\left[\left|\mathrm{lab}\mathopen{}\left(X\right)\mathclose{}\right|\right]\mathclose{}$, where $\delta_\cdot\mathopen{}\left[\cdot\right]\mathclose{}$ is the Kronecker delta, $\mathrm{lab}\mathopen{}\left(\bm{x},l\right)\mathclose{}=l$ denotes the projection of the label-augmented state space $\mathcal{X}\times\mathcal{L}$ onto its discrete label space $\mathcal{L}$, and $\mathrm{lab}\mathopen{}\left(X\right)\mathclose{}=\left\{l:\left(\bm{x},l\right)\in X\right\}$.

The \ac{pdf} for a $\delta$-\ac{glmb} \ac{rfs} may be parameterized by components $\left(I,h\right)\in\mathcal{F}\mathopen{}\left(\mathcal{L}\right)\mathclose{}\times\mathcal{H}$, where $\mathcal{H}$ is a discrete space, and associated weights $w^{\left(I,\,h\right)}$.
For \ac{mtt}, each component typically represents a data association hypothesis, with $I$ being a set of objects that may exist, $h$ being their combined measurement association history, and $w^{\left(I,\,h\right)}$ being the
estimated probability that the hypothesis is true.
A $\delta$-\ac{glmb} \ac{rfs} density is of the form
\begin{equation}
\pi\mathopen{}\left(X\right)\mathclose{}=\Delta\mathopen{}\left(X\right)\mathclose{}\sum_{\left(I,\,h\right)\in\mathcal{F}\mathopen{}\left(\mathcal{L}\right)\mathclose{}\times\mathcal{H}}w^{\left(I,\,h\right)}\delta_{I}\mathopen{}\left[\mathrm{lab}\mathopen{}\left(X\right)\mathclose{}\right]\mathclose{}\left(p^{\left(h\right)}\right)^{X}\,,\label{e:deltapdf}
\end{equation}
where $f^{X}=\prod_{\bm{x}\in X}f\mathopen{}\left(\bm{x}\right)\mathclose{}$ is the multi-object exponential and $p^{\left(h\right)}\mathopen{}\left(\cdot,l\right)\mathclose{}$ is an object's state \ac{pdf} given association history $h$ and label $l$.

Given the initial filtering density in Eq.~\ref{e:deltapdf} at time step $j$, the
predicted and updated density at time step $j+1$ given measurement set $Z_+$ is
\begin{equation}
\pi_+\mathopen{}\left(X_+\right)\mathclose{}\propto\Delta\mathopen{}\left(X_+\right)\mathclose{}\sum_{I,\,h,\,I_+,\,\theta_+}w^{\left(I,\,h\right)}w_+^{\left(I,\,h,\,I_+,\,\theta_+\right)}\mathopen{}\left(Z_+\right)\mathclose{}\delta_{I_+}\mathopen{}\left[\mathrm{lab}\mathopen{}\left(X_+\right)\mathclose{}\right]\mathclose{}\left(p_+^{\left(h,\,\theta_+\right)}\mathopen{}\left(\cdot\;\middle|\;Z_+\right)\mathclose{}\right)^{X_+}\,,\label{e:jpu}
\end{equation}
where $I\in\mathcal{F}\mathopen{}\left(\mathcal{L}\right)\mathclose{}$, $h\in\mathcal{H}$,
$I_+\in\mathcal{F}\mathopen{}\left(\mathcal{L}_+\right)\mathclose{}$, $\theta_+\in\Theta_+$,
$\mathcal{L}_+=\mathcal{L}\cup\mathcal{B}_+$, $\mathcal{B}_+$ is the space of object
labels that could be born at this time, $\Theta_+$ is the set of maps
$\theta_+:\mathcal{L}_+\rightarrow\left\{0:\left|Z_+\right|\right\}$ assigning
measurements in $Z_+$ to object labels, where $\theta_+\mathopen{}\left(l\right)\mathclose{}=0$ implies that label $l$ is not assigned a measurement,
and $w_+^{\left(I,\,h,\,I_+,\,\theta_+\right)}\mathopen{}\left(Z_+\right)\mathclose{}$ is the weight of the new hypothesis $\left(I,h,I_+,\theta_+\right)$ given prior hypothesis $\left(I,h\right)$ and measurement set $Z_+$.
For the full definition of these terms, as well as implementation details, see \cite{Vo2017}.

The number of possible hypotheses grows exponentially over time.
To maintain computational tractability, the set of new hypotheses
$\left(I,h,I_+,\theta_+\right)$ in Eq.~\ref{e:jpu} resulting from each prior
hypothesis $\left(I,h\right)$ is truncated using a ranked assignment algorithm and a cost matrix based on the terms that make up $w_+^{\left(I,\,h,\,I_+,\,\theta_+\right)}\mathopen{}\left(Z_+\right)\mathclose{}$.
The classic approach to solve this ranked assignment problem is to use Murty's
algorithm \cite{murty1968,Vo2013,Vo2014}, but for large cost matrices, a Gibbs
sampler-based approach is more efficient \cite{Vo2017}.

\section{Adaptive Multi-Fidelity Model Selection} \label{sec:methods}

The accuracy and computational efficiency of the multi-fidelity solution depend on the choice of low- and high-fidelity models. We seek the optimal model pair that minimizes computation time while satisfying a user-defined requirement for accuracy in the multi-fidelity solution. Accordingly, the low- and high-fidelity force models must incorporate sufficient perturbing forces to maintain the desired accuracy while avoiding extraneous perturbations that would unnecessarily increase computational cost. Thus, for cislunar applications, the optimal model pair will vary throughout cislunar space as the relative contributions of Earth gravity, Lunar gravity, and perturbing forces change as a function of position.

For the filtering applications considered in this work, we would ideally define the accuracy requirement as an upper limit on the \ac{pdf} divergence between the multi-fidelity and a full-fidelity solution in the prediction step of the filter, as proposed by~\cite{ben_amos}. Formulated as an optimization problem, this becomes
\begin{equation} \label{eq:pdf_optimization_problem}
\begin{array}{ll}
\min\limits_{\left(\bm{f}^L,\,\bm{f}^H\right)\in F\times F} & t_R\left(\bm{f}^L,\bm{f}^H\right) \\[6pt]
\text{subject to} & \mathcal{D}\left(p, \hat{p}\left(\bm{f}^L,\bm{f}^H\right)\right) < \varepsilon_\mathcal{D},
\end{array}
\end{equation}
where $\bm{f}^L$ and $\bm{f}^H$ are the low- and high-fidelity models, respectively, that solve the optimization problem and belong to a set of force models $F = \{ \bm{f}_1, \bm{f}_2, \dots \}$ that vary in which perturbing forces are included. We seek a map
\begin{equation}\label{eq:map}
    m: \mathcal{X} \to F \times F \quad \text{such that} \quad (\bm{f}^L,\bm{f}^H) = m\left(\{\bm{x}(t_j,\bm{\xi}_i)\}_{i=1}^m\right),
\end{equation}
where $\mathcal{X}$ is the state-space and $\{\bm{x}(t_j,\bm{\xi}_i)\}_{i=1}^m$ is the set of particles approximating the state-space \ac{pdf} (i.e., Monte Carlo samples) at some time $t_j$. Also, $t_R(\bm{f}^L,\bm{f}^H)$ is the runtime of selecting and evaluating the model pair $(\bm{f}^L,\bm{f}^H)$, $\mathcal{D}(\mathord{\cdot},\mathord{\cdot})$ is a \ac{pdf} divergence metric, $p$ is the state-space \ac{pdf} predicted from time $t_j$ to $t_{j+1}$ using a full-fidelity model, $\hat{p}(\bm{f}^L,\bm{f}^H)$ is the \ac{pdf} predicted using the multi-fidelity algorithm with $\bm{f}^L$ and $\bm{f}^H$, and $\varepsilon_\mathcal{D}$ is a user-defined value.

As posed, this optimization problem is intractable to compute online because evaluating $\mathcal{D}(p, \hat{p}(\bm{f}^L,\bm{f}^H))$ requires both a full-fidelity solution for $p$, which is prohibitively expensive to compute, and the multi-fidelity solution $\hat{p}\left(\bm{f}^L,\bm{f}^H\right)$, which is itself the unknown quantity we seek as the end result, making its direct computation in the optimization problem impractical. We therefore reformulate the optimization problem to have an accuracy requirement that the expected error in acceleration magnitude remain below some threshold. Since evaluating the acceleration at a particular state $\bm{x}(t_j,\bm{\xi}_i)$ does not require the numerical solution to the differential equations of the force models, this makes the problem formulation tractable. Moreover, by enforcing a bound on the mean-square acceleration error of the force model, we indirectly enforce mean-square convergence of the resulting state trajectories under standard regularity assumptions of the dynamics. Since mean-square convergence implies convergence in distribution, this acceleration-based criterion is, in that sense, stronger than the original requirement of convergence in density of the predicted \acp{pdf}.

Additionally, we will simplify the optimization problem by considering only one low-fidelity model over the space $\mathcal{X}$ (over the entire cislunar domain). This not only eliminates a decision variable, but also allows us to use the trajectory from the low-fidelity propagation to inform the model choice for the high-fidelity propagation. This is possible within the multi-fidelity framework since the low-fidelity propagation is always done before the high-fidelity propagation, meaning the high-fidelity model does not yet need to be determined when the low-fidelity propagation is performed. Moreover, a truncated force model that is constant across cislunar space may still account for the dominant forces, and as demonstrated by the numerical simulations in Section \ref{sec:numeric_simulations}, yields sufficient multi-fidelity accuracy and runtime, making it justified from a dynamical and computational standpoint.

The reformulated optimization problem is then
\begin{equation} \label{eq:accel_optimization_problem}
\begin{array}{rl}
\min\limits_{\bm{f}^H\in F} \quad & t_R\left(\bm{f}^L,\bm{f}^H\right) \\[8pt]
\text{subject to} \quad & \mathbb{E} \left[ a_{\text{error}} \right] < \varepsilon_{a},
\end{array}
\end{equation}
where $\mathbb{E} \left[ a_{\text{error}} \right]$ is the expected value of the error in acceleration magnitude and $\varepsilon_{a}$ is a user-defined value. The high-fidelity force model that solves this optimization problem is the one that includes only the minimal set of perturbations necessary to satisfy the accuracy requirement for a given region of cislunar space, as incorporating additional perturbations generally increases runtime. Since the high-fidelity model is selected as a function of position in cislunar space, only perturbations that vary spatially need to be adjusted -- specifically, the non-spherical gravitational perturbations of the Earth and Moon (which we note also vary temporally due to the rotation of these bodies). As a space object approaches one of these bodies, its non-spherical gravity terms become more significant and should be included in the model. Conversely, as the object moves farther away, those terms diminish in influence and can be truncated.

Among orbital perturbations, non-spherical gravity is one of the primary contributors to computational cost, as evaluating spherical harmonic expansions becomes increasingly expensive with degree and order; in particular, the number of terms (and thus the computational complexity) grows quadratically with the maximum degree and order~\cite{McArdle}, and these evaluations must be performed at every integration step. In contrast, perturbations such as \ac{srp} and the Sun’s gravity remain relatively constant across cislunar space and impose negligible computational overhead compared to spherical harmonic evaluations. Therefore, these perturbations are kept constant in both the low- and high-fidelity models. Additionally, Earth atmospheric drag is not considered, as trajectories that encounter significant atmospheric effects are better suited to an Earth-centric multi-fidelity framework rather than a cislunar one. The low- and high-fidelity models $\bm{f}^L$ and $\bm{f}^H$ for $\bm{f}$ in Eq.~\ref{eq:dyn_sys} are then
\begin{equation}\label{eq:lf_force_model}
    \bm{f}^L = \begin{bmatrix}
        \dot{\bm{r}} \\
        \bm{a}_{2B} + \sum_{n} \bm{a}_{NB,n} + \bm{a}_{SR}
    \end{bmatrix},
\end{equation}
\begin{equation}\label{eq:hf_force_model}
    \bm{f}^H = \begin{bmatrix}
        \dot{\bm{r}} \\
        \bm{a}_{2B} + \sum_{n} \bm{a}_{NB,n} + \sum_{i} \bm{a}_{SH,i} + \bm{a}_{SR}
    \end{bmatrix},
\end{equation}
where $\bm{a}_{2B}$ is the acceleration due to the central body (the two-body term), $\bm{a}_{NB,n}$ is the acceleration due to other celestial bodies (the $n$-body term), $\bm{a}_{SH,i}$ is the acceleration due to spherical harmonic gravity on the $i$th body, and $\bm{a}_{SR}$ is the acceleration from \ac{srp}, modeled with a ``cannonball'' model assuming constant area and reflectance and a simple eclipsing function. The equations for these accelerations are provided by \cite{TurboProp_manual}.

Since non-spherical gravity is the only perturbation that varies in the high-fidelity model, we focus on determining the appropriate spherical harmonic expansion degree and order as a function of position in space and acceptable acceleration error. The solution is obtained separately for each celestial body's gravity (in this work, the Earth and Moon) using a lookup table from \cite{McArdle} and \cite{McArdle_GNC}, which provides the appropriate expansion degree and order for a given altitude to a celestial body and user-defined acceleration noise, thereby defining the mapping in Eq.~\ref{eq:map}.

For a spherical harmonic expansion of degree and order $L$, \cite{McArdle} defines the total acceleration error as
\begin{equation} \label{eq:omission_commission_error}
    \mathbb{E}[\|\nabla U_{T,L} \|^2] = 
    \mathbb{E}[\|\nabla U_{O,L} \|^2] + 
    \mathbb{E}[\|\nabla U_{C,L} \|^2],
\end{equation}
where $\mathbb{E}[\|\nabla U_{O,L} \|^2]$ is the omission error, $\mathbb{E}[\|\nabla U_{C,L} \|^2]$ is the commission error, and $U$ is the gravitational potential. The omission error represents the expected acceleration error due to truncation of spherical harmonic terms beyond degree and order $L$, while the commission error accounts for the expected acceleration error due to published uncertainties in the included spherical harmonic terms up to and including degree and order $L$. The mean omission and commission errors are computed by averaging the squared acceleration errors in Cartesian coordinates for the celestial body in question.

The $x$-, $y$-, and $z$-components of error are assumed to be Gaussian distributed, meaning the squared error in Eq.~\ref{eq:omission_commission_error} follows a chi-squared distribution. Reference~\cite{McArdle} then computes the upper bound of a 99.7\% confidence interval for all expansion degrees to produce the lookup table. Our adaptive high-fidelity model selection method uses the low-fidelity trajectory to determine the space object's altitude relative to the Earth and Moon at discrete points along the trajectory. It then queries the appropriate lookup table to select the expansion degree and order to use in the high-fidelity model for the current position in cislunar space.\footnote{This work uses a C++ port of the MATLAB code published in \cite{McArdle} (available at \url{https://doi.org/10.5281/zenodo.6814713}, accessed January 2025).}

In regions of cislunar space far from either body, the recommended expansion degree and order for both the Earth and Moon may approach zero. When this occurs, the low- and high-fidelity models become sufficiently similar such that no multi-fidelity correction is needed. Therefore, at a specified threshold for the high-fidelity model's expansion degree, our implementation performs only a low-fidelity propagation and skips the multi-fidelity procedure. The sequence of steps for the adaptive multi-fidelity method is summarized in Algorithm~\ref{alg:mf_prop}.

\begin{algorithm}
\caption{Adaptive Multi-Fidelity Propagation}\label{alg:mf_prop}
\begin{algorithmic}[1]
\Statex \textbf{Input:} Realizations of random inputs (Monte Carlo samples) $\Xi=\{\bm{\xi}_i\}_{i=1}^m$, maximum acceleration error $\varepsilon_{a}$, expansion degree/order $L_{\text{min}}$ below which to skip multi-fidelity correction
\Statex \textbf{Output:} Multi-fidelity samples $\{\hat{\tilde{\bm{x}}}^H(\bm{\xi}_i)\}_{i=1}^m$
\State Propagate all particles with the low-fidelity model (Eq.~\ref{eq:lf_force_model}) to get $\bm{X}^L(\Xi)$
\State Compute altitude and corresponding gravity expansion degree and order $L$ for the Earth and Moon at discrete intervals along the low-fidelity trajectories using the lookup table
\If{$L > L_{\text{min}}$}
    \State Identify important samples $\bar{\Xi}$ with Eqs.~\ref{eq:argmax}--\ref{eq:imp_samp_union}
    \State Propagate $\bar{\Xi}$ with the high-fidelity model (Eq.~\ref{eq:hf_force_model}), with $L$ changing at each interval used in step~2
    \State Compute the multi-fidelity solution with Eqs.~\ref{eq:hf_expansion} and \ref{eq:LL^Tc}
\Else
    \State Skip multi-fidelity procedure, output $\bm{X}^L(\Xi)$
\EndIf
\end{algorithmic}
\end{algorithm}

To obtain the time steps and associated state vectors used in the multi-fidelity algorithm's snapshot matrix (Eq.~\ref{eq:snapshot}), we propagate the particles backward in time $N$ time steps rather than store their histories to reduce memory footprint, as was done in \cite{ben_amos}. This is beneficial in the context of \ac{mtt} since storing the time histories for multiple different tracks can become overly memory intensive. For applications where speed is desired over memory, the time histories could instead be stored from the initial forward propagation, rather than doing the backward propagation.

Lastly, the number of time steps serves as a user-defined tuning parameter. Since the rank of the snapshot matrix limits the number of important samples that can be identified, increasing the number of time steps generally raises the matrix rank, enabling the identification of more important samples and enhancing the accuracy of the multi-fidelity solution. However, as pointed out in \cite{petion_mf_ar}, this comes at the cost of increased computation, as additional important samples must be propagated, and the improvement in accuracy exhibits diminishing returns. Eventually, the state vectors introduced from additional time steps lack sufficient linear independence to increase the rank of the snapshot matrix. In general, longer propagation times warrant the use of more time steps.
\section{Numeric Simulations} \label{sec:numeric_simulations}
\subsection{Test Case Descriptions} \label{sec:test_case_descriptions}

This section provides the test case scenarios and parameters that are used for assessing the adaptive multi-fidelity model selection method in the next two subsections. Five scenarios are used: a \ac{dro}, a \ac{nrho}, a \ac{lto}, a low Lunar flyby, and a \ac{llo}, with the first four scenarios using orbital parameters from \cite{ben_amos}. Orbits such as these are candidates for upcoming cislunar missions, and are therefore relevant to cislunar \ac{ssa}. Furthermore, these orbits span large swaths of cislunar space, making them challenging test cases for the adaptive method.

\begin{figure}[htbp]
    \centering
    \includegraphics[scale = 0.67]{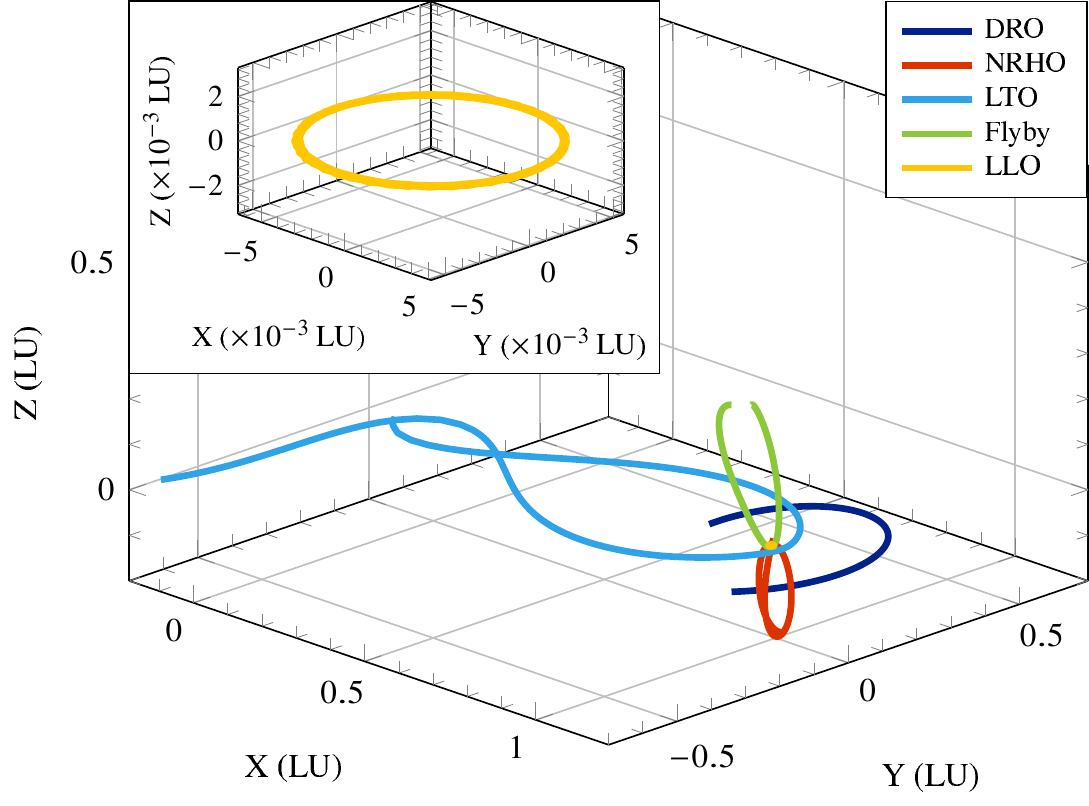}
    \caption{Nominal trajectories for each cluster in the rotating barycentric frame.
    Inset shows the LLO in Moon-centered coordinates in the rotating frame.}
    \label{fig:mean_trajectories}
\end{figure}

In the tracking results presented in Section~\ref{sec:results_tracking}, each of the five scenarios includes a cluster of five \acp{aso}. The \acp{aso} in each cluster are sampled from the same initial state \ac{pdf}, with the means and variances provided in Table~\ref{tab:cluster_means} and diagonal covariances of $\sigma^2_{\text{state}}\bm{I}_6$. Figure~\ref{fig:mean_trajectories} depicts the nominal trajectory for each of the five scenarios, computed from a deterministic propagation of the mean for ten days for the \ac{dro}, \ac{nrho}, \ac{lto}, and flyby (the length of these scenarios), and \SI{24}{\hour} for the \ac{llo} (the length of the \ac{llo} scenario). The nominal trajectory of the Lunar flyby scenario has a close approach altitude with the Moon of \SI{133}{\km}, assuming a spherical Moon. The nominal \ac{llo} trajectory has periapsis altitude \SI{125}{\km}, again assuming a spherical Moon, and period \SI{143}{\minute}. However, due to random realizations of the initial conditions, the closest approaches in the test cases will vary.
\begin{table}[htbp]
    \centering
    \footnotesize
    \renewcommand{\arraystretch}{1.25} 
    \setlength{\tabcolsep}{4pt} 
    \caption{Initial means in normalized units (LU, LU/TU) and variances (LU\textsuperscript{2}, LU\textsuperscript{2}/TU\textsuperscript{2}) for each scenario, expressed in the rotating barycentric frame.}
    \begin{tabular}{c r r r r r r r}
        \toprule
        Scenario & \multicolumn{6}{c}{Mean} & $\sigma^2_{\text{state}}$ \\
        & $x$ & $y$ & $z$ & $\dot{x}$ & $\dot{y}$ & $\dot{z}$ \\
        \midrule
        \ac{dro}  & 0.806  & 0.000  & 0.000  & 0.000  & 0.519  & 0.000 & $10^{-4}$  \\
        \ac{nrho} & 1.022  & 0.000  & -0.182 & 0.000  & -0.103 & 0.000 & $10^{-4}$  \\
        \ac{lto}  & -0.112 & 0.000  & 0.000  & 2.194  & -3.440 & 0.000 & $10^{-4}$  \\
        Flyby     & 0.949  & -0.019 & 0.304  & -0.006 & 0.064  & 0.003 & $10^{-4}$  \\
        \ac{llo}  & 0.993  & 0.000  & 0.000  & 0.000  & 1.570  & 0.000 & $10^{-5}$  \\
        \bottomrule
    \end{tabular}
    \label{tab:cluster_means}
\end{table}

The acceleration noise limit from Eq.~\ref{eq:accel_optimization_problem} is set to $\varepsilon_a=\SI{e-15}{\km\per\s^2}$ for all simulations. This value was found to provide a suitable balance between increased computation time from additional spherical harmonic gravity terms and accuracy in propagated states.

The low-fidelity, high-fidelity, and ``truth'' propagators all use a 4th order accurate Runge-Kutta integrator with a 5th order error estimate and adaptive step size~\cite{dormand_prince}. The low-fidelity model uses Earth and Moon point-mass gravity and the ``truth'' model uses 120$\times$120 gravity for both bodies. The degree and order for the adaptive method never exceeds that of the truth in these test cases. The Earth and Moon use the EGM2008~\cite{egm2008} and LP165P~\cite{LP165P} gravity fields, respectively, as well as the IAU2006 coordinate system reduction~\cite{IAU} for the Earth. Models of all fidelity have cannonball \ac{srp} and Sun point-mass gravity. Additional force model parameters from \cite{jones_weisman} may be found in Table~\ref{tab:satellite_parameters}. The positions of celestial bodies are determined by the JPL DE440 ephemerides~\cite{jpl_ephem}.

\begin{table}[htbp]
\centering
\small
\renewcommand{\arraystretch}{1.25} 
\setlength{\tabcolsep}{4pt} 
\caption{Force model and satellite parameters.}
\begin{tabular}{l l}
\toprule
Parameter & Value                                      \\ \midrule
Satellite mass                               & 500 kg                    \\
Satellite SRP area                           & 1 m\textsuperscript{2}    \\
Reflectivity coefficient                     & 1.5                       \\
Epoch time                                   & 2455200.5 UTC             \\
\bottomrule
\end{tabular}
\label{tab:satellite_parameters}
\end{table}

Runtime tests are generated on a Dell Precision Tower 3430 desktop computer running Red Hat Enterprise Linux 8.10 and the Linux 4.18.0 kernel with a 3.2 GHz Intel Core i7-8700 processor and 16 GB of random-access memory. The propagation software is written in C and all other software (tracker and adaptive model selection) is written in C++ and compiled with the GCC 8.5.0 compiler.

\subsection{Results: Multi-Fidelity Propagation} \label{sec:results_mf_propagation}

This section evaluates the performance of the adaptive multi-fidelity method solely in the context of propagation, without incorporating it into a tracking framework. Analyzing the propagation in isolation is informative due to its broad applicability to uncertainty propagation. Here, a particle ensemble is propagated using a Monte Carlo approach, though the multi-fidelity method can also be applied to Gaussian mixture propagation via the unscented transform~\cite{jones_weisman}. Additionally, this analysis can be viewed as the prediction step of a filtering process. The integration of the method into a full filtering and tracking framework is presented in the subsequent section.

Process noise is excluded from the propagation in this section to isolate the accuracy of the adaptive multi-fidelity method from the confounding effects that stochastic processes may introduce. The degree and order of the spherical harmonic gravity expansion is updated in time increments $\Delta t$, where $\Delta t$ is \SI{10}{\minute} in the \ac{llo} case and \SI{1}{\hour} otherwise, and the multi-fidelity snapshot matrix is constructed using seven time steps in increments of $\Delta t$ at the end of the propagation. If a dense-output ODE solver is used, states at intermediate time steps for Step~2 of Algorithm~\ref{alg:mf_prop} may be computed with negligible additional computational cost. If an ODE solver without dense output is used (as is the case with \cite{TurboProp_manual}, which was used to generate these results), solving for state values at many small time steps may increase computation time. Therefore, we only request state values from the integrator at the ends of the $\Delta t$ time intervals when above \SI{2000}{\km} altitude, and in $\frac{1}{4}\Delta t$ intervals below \SI{2000}{\km} altitude where the dynamics evolve more quickly. All test cases use 1000 low-fidelity samples with the number of high-fidelity important samples being the rank of the snapshot matrix in Eq.~\ref{eq:snapshot}, as suggested by~\cite{petion_mf_ar}.

To evaluate the adaptive multi-fidelity method, four non-adaptive multi-fidelity baselines are considered, each using a constant gravity expansion for the high-fidelity model throughout the entire propagation: 30$\times$30, 60$\times$60, 90$\times$90, and the maximum degree and order found by the adaptive method for the particular scenario. In a non-adaptive approach, an operator tasked with propagating uncertainty or tracking space objects may not have prior knowledge of the specific regions of cislunar space through which the objects will pass, and therefore would have to select the gravity expansion heuristically. The 30$\times$30, 60$\times$60, and 90$\times$90 non-adaptive methods are chosen to represent the range of possible gravity expansions that might be used in the absence of such prior knowledge and fall within the set of gravity expansions used by the adaptive method in Secs.~\ref{sec:results_mf_propagation} and \ref{sec:results_tracking}. All other aspects of the force models in the non-adaptive method are identical to the force models in the adaptive method. For the propagation results in this section, the multi-fidelity correction is never skipped (i.e., $L_{\text{min}}$ in Algorithm~\ref{alg:mf_prop} is set to 0).

\begin{figure}[htbp]
    \centering
    \includegraphics[scale = 0.53]{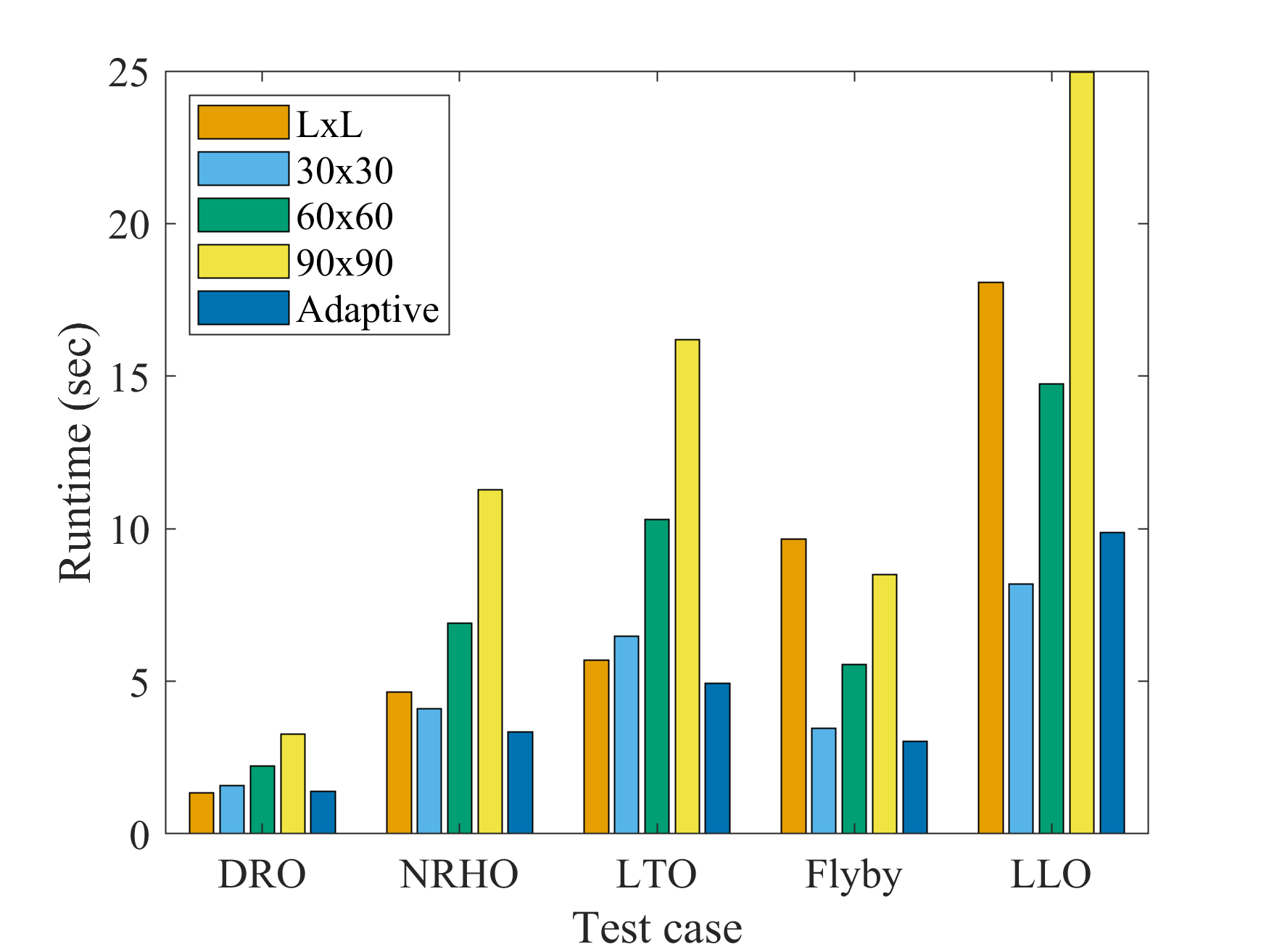}
    \caption{Runtime of non-adaptive and adaptive multi-fidelity methods for all test cases.}
    \label{fig:prop_runtime}
\end{figure}

\begin{figure}[htbp]
    \centering
    \includegraphics[scale = 0.53]{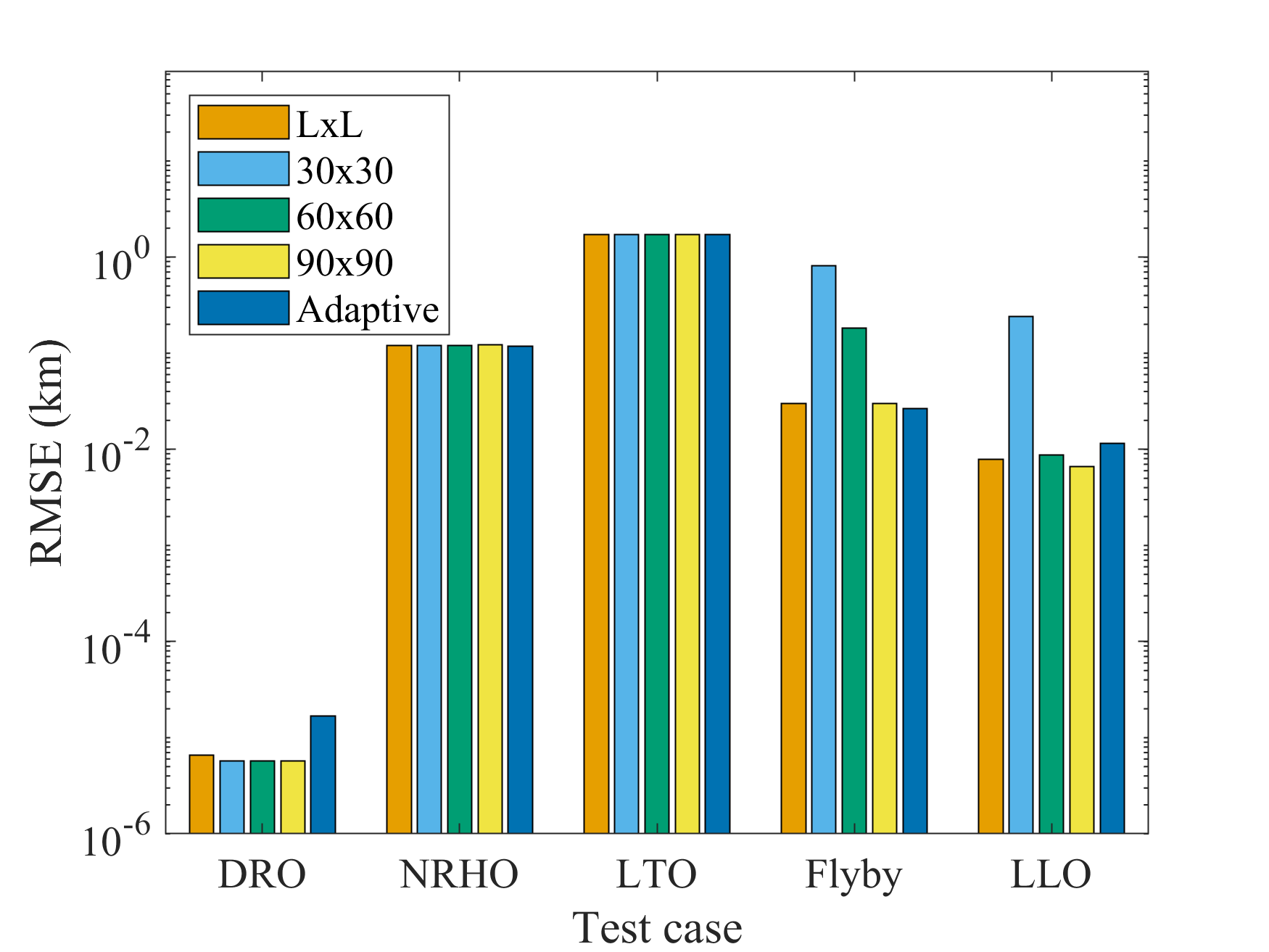}
    \caption{\ac{rmse} of non-adaptive and adaptive multi-fidelity methods for all test cases. Note the logarithmic scaling of the vertical axis.}
    \label{fig:prop_accuracy}
\end{figure}

Runtime and accuracy statistics may be found in Tables~\ref{tab:prop_runtime} and \ref{tab:prop_accuracy}, respectively. Each set of results reflects the average of 10 Monte Carlo simulations, with each simulation having a different set of random initial states sampled from the distribution described for each test case in Section~\ref{sec:test_case_descriptions}. For each Monte Carlo simulation, the same set of initial states is used for all propagation methods (low-fidelity, high-fidelity, non-adaptive multi-fidelity, and adaptive multi-fidelity). The position \ac{rmse} is computed for each propagation method using the “truth” defined in Section~\ref{sec:test_case_descriptions}. In Tables~\ref{tab:prop_runtime} and \ref{tab:prop_accuracy} and Fig.~\ref{fig:prop_runtime} and \ref{fig:prop_accuracy}, $L \times L$ indicates the maximum degree and order found by the adaptive method for that particular test case across all ten Monte Carlo runs, and is reported in the last row of each of Tables~\ref{tab:prop_runtime} and \ref{tab:prop_accuracy}. In these tables, the low-fidelity results reflect using only the low-fidelity model to propagate the entire particle ensemble. Likewise, the high-fidelity results reflect using only the high-fidelity model to propagate all particles, with the results presented for all four (30$\times$30, 60$\times$60, 90$\times$90, and $L \times L$) non-adaptive high-fidelity models. The non-adaptive multi-fidelity results reflect using those four non-adaptive high-fidelity models in the multi-fidelity framework. The non-adaptive and adaptive multi-fidelity runtime and accuracy results are plotted in Fig.~\ref{fig:prop_runtime} and \ref{fig:prop_accuracy}, respectively.

\begin{table}[htbp]
    \centering
    \footnotesize
    \renewcommand{\arraystretch}{1.25} 
    \setlength{\tabcolsep}{4pt} 
    \caption{Runtime (sec) for all test cases. LF, HF, and MF are shorthand for low-fidelity, high-fidelity, and multi-fidelity, respectively.}
    \begin{tabular}{l r r r r r r}
        \toprule
        Method & \multicolumn{5}{c}{Scenario} \\
        & {\ac{dro}}   & {\ac{nrho}}   & {\ac{lto}}   & {Flyby}   & {\ac{llo}}     \\
        \midrule
        LF                   & \num{1.054}   & \num{2.102}   & \num{3.637}   & \num{1.879}   & \num{3.866}   \\
        $L \times L$ HF      & \num{11.82}   & \num{73.41}   & \num{73.65}   & \num{237.8}   & \num{353.5}   \\
        30$\times$30 HF      & \num{23.29}    & \num{56.15}   & \num{101.5}   & \num{49.00}   & \num{110.7}   \\
        60$\times$60 HF      & \num{55.82}   & \num{140.6}   & \num{244.9}   & \num{117.1}   & \num{268.9}   \\
        90$\times$90 HF      & \num{108.2}   & \num{273.9}    & \num{476.5}   & \num{217.1}   & \num{537.5}   \\
        $L \times L$ MF      & \num{1.352}   & \num{4.650}   & \num{5.692}     & \num{9.669}     & \num{18.08}   \\
        30$\times$30 MF      & \num{1.592}   & \num{4.108}   & \num{6.477}   & \num{3.453}   & \num{8.184}   \\
        60$\times$60 MF      & \num{2.231}   & \num{6.894}   & \num{10.31}   & \num{5.555}   & \num{14.75}   \\
        90$\times$90 MF      & \num{3.262}   & \num{11.28}   & \num{16.20}   & \num{8.499}   & \num{24.96}   \\
        Adaptive             & \num{1.390}   & \num{3.346}   & \num{4.922}   & \num{3.029}   & \num{9.885}   \\
        Truth                & \num{175.6}   & \num{430.4}   & \num{794.5}    & \num{378.4}   & \num{876.0}   \\
        \midrule
        $L \times L$ & 3 & 39 & 20 & 91 & 71 \\
        \bottomrule
    \end{tabular}
    \label{tab:prop_runtime}
\end{table}

\begin{table}[htbp]
    \centering
    \footnotesize
    \renewcommand{\arraystretch}{1.25} 
    \setlength{\tabcolsep}{4pt} 
    \caption{Position \ac{rmse} (km) for all test cases.}
    \begin{tabular}{l r r r r r r}
        \toprule
        Method & \multicolumn{5}{c}{Scenario} \\
        & {\ac{dro}}   & {\ac{nrho}}   & {\ac{lto}}   & {Flyby}   & {\ac{llo}}     \\
        \midrule
        LF                   & \num{0.6323}   & \num{203.4}    & \num{299.2}     & \num{42.86}     & \num{102.6}   \\
        $L \times L$ HF      & \num{4.770e-6} & \num{0.03239}  & \num{6.882e-6} & \num{0.006116} & \num{0.003720} \\
        30$\times$30 HF      & \num{0}        & \num{0.03296}  & \num{1.454e-9} & \num{0.8167}   & \num{0.2441}   \\
        60$\times$60 HF      & \num{0}        & \num{0.03195}  & \num{0}        & \num{0.1828}   & \num{0.007048} \\
        90$\times$90 HF      & \num{0}        & \num{0.02604}  & \num{0}        & \num{0.006414} & \num{0.001509} \\
        $L \times L$ MF      & \num{6.565e-6} & \num{0.1203}   & \num{1.726}    & \num{0.03008}  & \num{0.007872} \\
        30$\times$30 MF      & \num{5.758e-6} & \num{0.1201}   & \num{1.726}    & \num{0.8207}   & \num{0.2428}   \\
        60$\times$60 MF      & \num{5.758e-6} & \num{0.1206}   & \num{1.726}    & \num{0.1814}   & \num{0.008804} \\
        90$\times$90 MF      & \num{5.758e-6} & \num{0.1215}   & \num{1.726}    & \num{0.03021}  & \num{0.006661} \\
        Adaptive MF          & \num{1.684e-5} & \num{0.1188}   & \num{1.726}    & \num{0.02669}  & \num{0.01151}  \\
        \midrule
        $L \times L$ & 3 & 39 & 20 & 91 & 71 \\
        \bottomrule
    \end{tabular}
    \label{tab:prop_accuracy}
\end{table}

The multi-fidelity algorithm achieves a multiple-order-of-magnitude improvement in the \ac{rmse} compared to the low-fidelity solution across all test cases, yielding kilometer to sub-kilometer accuracy, depending on the scenario. Additionally, the use of multi-fidelity propagation significantly reduces runtime relative to high-fidelity propagation, demonstrating an approximately tenfold speedup across test cases. The difference between high- and multi-fidelity runtime is amplified as the degree and order of the gravity expansion used in the high-fidelity model increase. The \ac{rmse} also generally decreases with more terms in the gravity expansion, though Table~\ref{tab:prop_accuracy} indicates diminishing returns. Notably, for the \ac{dro} and \ac{lto} test cases, the high-fidelity propagation reaches zero \ac{rmse} relative to the truth model. This arises because these test cases remain at high altitudes, where, beyond a certain degree and order expansion, non-spherical gravitational accelerations fall below machine precision. This highlights a benefit of the adaptive method: by dynamically adjusting the gravity expansion based on an acceleration error threshold, it prevents the inclusion of unnecessary terms, thereby reducing computational cost.

As shown in Tables~\ref{tab:prop_runtime} and \ref{tab:prop_accuracy} and Fig.~\ref{fig:prop_runtime} and \ref{fig:prop_accuracy}, the adaptive multi-fidelity method consistently achieves nearly-equal or superior accuracy compared to the non-adaptive multi-fidelity approach while maintaining a lower runtime. This is most evident in the flyby and \ac{llo} test cases, where the non-adaptive method continues to improve in \ac{rmse} with increasing gravity expansion order but at the cost of greater runtime. In contrast, the adaptive method achieves the accuracy of the non-adaptive approach (from the $L \times L$ case) while maintaining a significantly faster runtime. For the flyby, this may be attributed to the orbit having the largest altitude variation of all test cases. At perilune, a large gravity expansion is necessary to maintain accuracy, but at higher altitudes during the remainder of the trajectory, these additional terms become extraneous. For the \ac{llo}, this may be attributed to the fact that at low altitudes, the required spherical harmonic degree and order for a given acceleration error is more sensitive to small changes in altitude. This indicates that the adaptive method is most beneficial for cases where the distance to the Earth or Moon varies significantly over the trajectory, or for low-altitude orbits in general.

\begin{figure}[htbp]
    \centering
    \includegraphics[scale = 0.44]{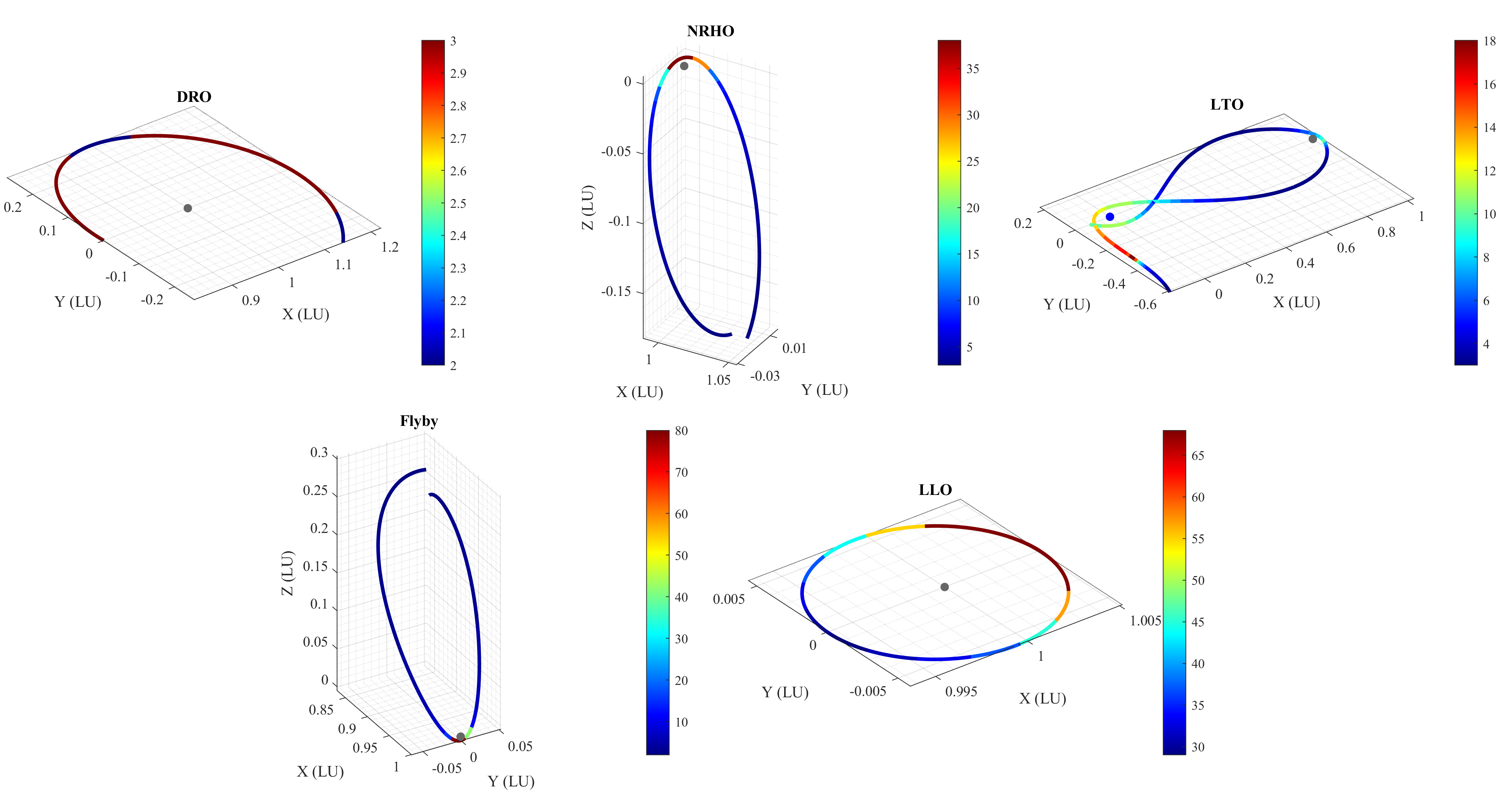}
    \caption{Mean trajectories colored by the spherical harmonic gravity degree and order of the Earth or Moon (whichever bodies' expansion is larger at the particular time step) for up to one revolution. Since the degree and order is computed across the entire particle set (not shown), the apsides of the mean trajectory may not correspond exactly with the extrema of degree and order. Color bar indicates degree and order. Note that the Earth (blue dot) and Moon (gray dot) are not shown to scale.}
    \label{fig:degord_trajectory}
\end{figure}

\begin{figure}[htbp]
    \centering
    \includegraphics[scale = 0.4]{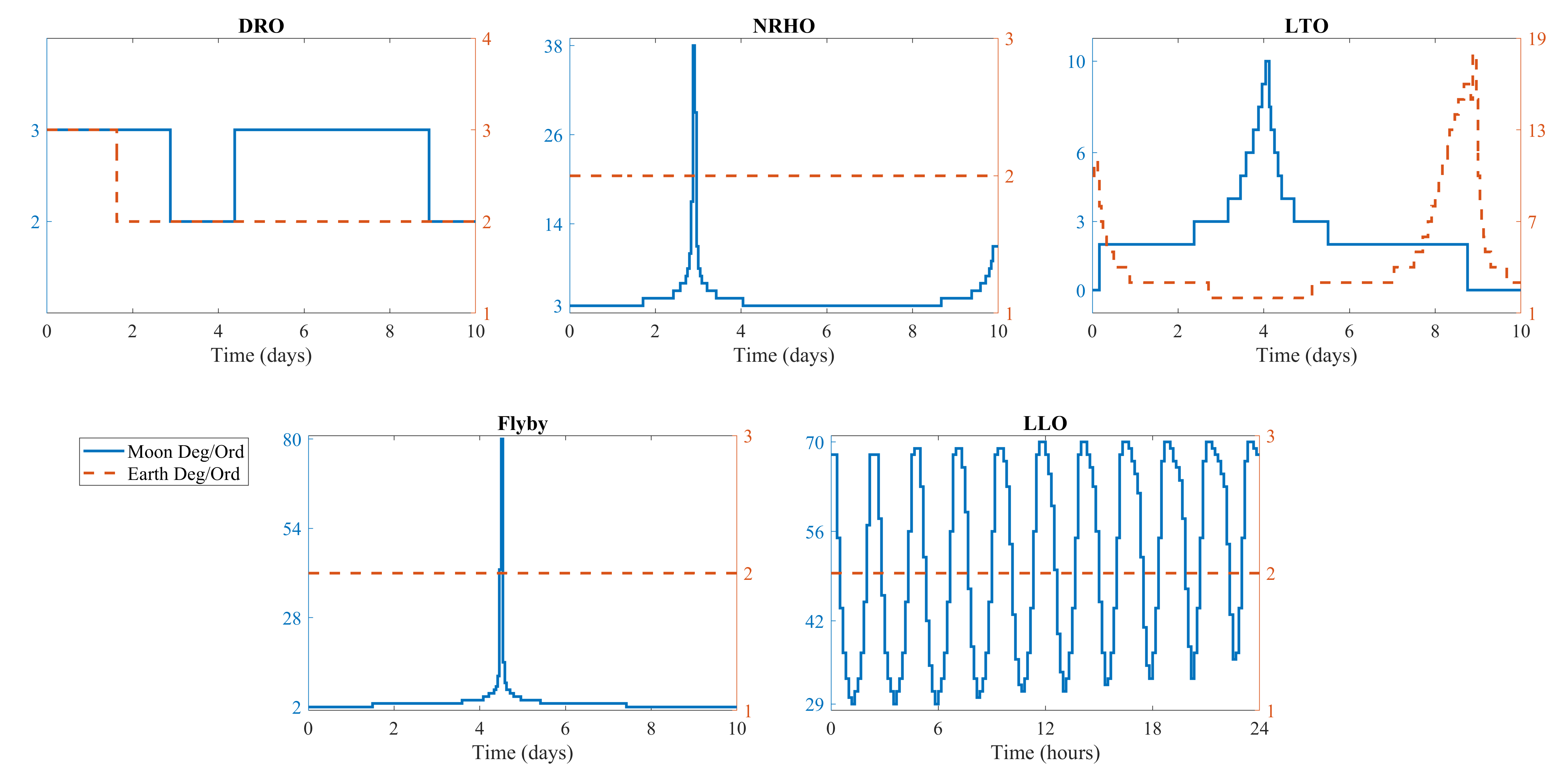}
    \caption{Spherical harmonic gravity degree and order as a function of time for the Moon (left vertical axis) and Earth (right vertical axis).}
    \label{fig:degord_time}
\end{figure}

Figure~\ref{fig:degord_trajectory} depicts the mean trajectory for each test case colored by the degree and order of the spherical harmonic gravity expansion. As expected, the degree and order increase closer to the Moon, reaching a maximum at perilune as the higher order terms gain significance. This effect is further demonstrated in Fig.~\ref{fig:degord_time}, which plots the degree and order as a function of time and exhibits a sharp peak at periapsis for cases that pass near the Earth or Moon. It should be observed that the degree and order in Fig.~\ref{fig:degord_trajectory} and \ref{fig:degord_time} is determined from the lowest altitude across the entire Monte Carlo particle set (not shown in the figure), as per Section~\ref{sec:methods}, and therefore the maximum and minimum degree and order may not correspond exactly with when the mean trajectory reaches periapsis and apoapsis. This is most apparent in the \ac{lto} test case, where some Monte Carlo samples stay near the Earth after the mean trajectory has passed perigee, causing the degree and order to remain high.

\subsection{Results: Multi-Target Tracking} \label{sec:results_tracking}
This section presents Monte Carlo analyses of the multi-target filter's performance in two simulated cislunar tracking scenarios.
Each scenario involves a single sensor in a polar lunar orbit tracking one or more clusters of space objects using the \ac{glmbf} with each of the following propagation methods: low-fidelity propagation with no correction, non-adaptive multi-fidelity propagation, and adaptive multi-fidelity propagation.
Each cluster contains five objects sampled from one of the initial state \acp{pdf} given in Table~\ref{tab:cluster_means}.
The first \ac{mtt} scenario combines the \ac{dro}, \ac{nrho}, \ac{lto}, and Lunar flyby scenarios from the previous subsection, with all twenty objects across
the four clusters being tracked simultaneously over a period of ten days.
In the second \ac{mtt} scenario, only the five-object \ac{llo} cluster is tracked over a period of \SI{24}{\hour}.
The \ac{llo} scenario is kept separate bcause a significantly higher observation rate is required to accurately track objects in \ac{llo} than the other orbit families.
This is implemented by setting the simulation time step $\Delta t$ to \SI{1}{\hour} in the combined scenario and \SI{10}{\minute} in the \ac{llo} scenario, as in the previous section.
Additionally, the high-fidelity model used in the non-adaptive multi-fidelity cases uses a higher degree and order in the \ac{llo} scenario to achieve comparable tracking accuracy to the adaptive method.

The low-fidelity, truth, and adaptive multi-fidelity propagators are the same as in the previous subsection, but when using the adaptive multi-fidelity method, the \ac{engmf} skips the multi-fidelity correction when the maximum recommended gravity model resolution has stayed less than 5$\times$5 since the last correction.
The non-adaptive multi-fidelity propagator in this subsection uses 30$\times$30 gravity field models for both bodies in the combined scenario and 70$\times$70 in the \ac{llo} scenario.
Note that these are below the values given in the last row of Table~\ref{tab:prop_runtime} for some orbits.
The \ac{engmf} uses \num{200} particles to parameterize each track.

These simulations use a single optical sensor located in a high Lunar polar orbit to obtain measurements of right ascension and declination and their respective rates.
The sensor's initial orbital state in normalized units in the rotating barycentric frame is $\left[0.988~~0~~0.018~~0~~0.788~~0\right]^\intercal$.
Sensor parameters from \cite{ben_amos} are used, with the angle and angle-rate measurements possessing Gaussian distributed errors with a mean of zero and standard deviation of \SI{0.1}{\arcsec} for the angles and \SI{0.001}{\arcsec\per\s} for the angle-rates.
The measurement errors are assumed to be uncorrelated.
The sensor generates measurements every 12 time steps ($12\Delta t$) after the start of the simulation, with probability of detection $P_D=\num{0.95}$.
Target occlusion by the Earth and Moon is not simulated.
The simulated sensor does not generate clutter returns and the multi-target filter assumes a constant clutter intensity of $\kappa=\num{e-9}$.

The multi-target filter is initialized with the correct number of objects with each of the initial state \acp{pdf} from Table~\ref{tab:cluster_means} that are included in the scenario under consideration.
Birth is not modeled in the filter, which assumes a survival probability near one for all objects, discards hypotheses with weights below $\num{e-3}$, and employs Murty’s algorithm to cap the number of hypotheses at \num{50}.
This limit is distributed among prior \ac{glmb} components in proportion to the square root of their respective weights.
The snapshot matrix is again constructed using the last seven time steps at the time of multi-fidelity correction.
Additionally, the \ac{engmf} assumes a process noise covariance $\bm{Q}=\sigma_a^2\bm{I}_3$, where the acceleration standard deviation is $\sigma_a=\SI{e-12}{\km\per\s^2}$.

The average results over 20 runs of each of the two scenarios with each of the three propagation methods are summarized in Table~\ref{t:mttres}.
The times shown are the total filtering time for one run of the simulation, broken down between track prediction (prediction time) and the rest of the \ac{glmbf} algorithm (update time).
Note that in the multi-fidelity cases, multi-fidelity correction (i.e., steps~2--8 of Algorithm~\ref{alg:mf_prop}) is included in the update time, because it is only performed when needed to facilitate a measurement update.
Tracking errors are averaged over the simulation time span.
Tracking error is defined here as the \ac{ospa} multi-object distance between the \ac{glmbf}'s estimate of the states of the objects and their true states \cite{schuhmacher2008}.
The \ac{ospa} metric used here is based on the Euclidean distance in Cartesian position space and has order \num{2} and cutoff \SI{100}{\km}.

\begin{table}[htbp]
\centering
\footnotesize
\renewcommand{\arraystretch}{1.25} 
\setlength{\tabcolsep}{4pt} 
\caption{Summary statistics for multi-target tracking simulations.
    Times are total values for one run of the simulation and tracking errors are averaged over the simulation time span.
    Update time includes multi-fidelity correction when applicable.}\label{t:mttres}
\begin{tabular}{lrrrrrr}
\toprule
\multicolumn{1}{c}{Method}&\multicolumn{3}{c}{Combined Scenario}&\multicolumn{3}{c}{LLO Scenario}\\
&Prediction Time (s)&Update Time (s)&Error (km)&Prediction Time (s)&Update Time (s)&Error (km)\\
\midrule
LF             &\num{14.965}& \num{9.623}&\num{20.241}&\num{11.620}&  \num{1.506}&\num{30.296}\\
Non-Adaptive MF&\num{15.860}&\num{59.603}&\num{11.832}&\num{12.201}&\num{181.093}& \num{5.210}\\
Adaptive MF    &\num{16.047}&\num{30.828}&\num{11.865}&\num{15.779}& \num{76.706}& \num{5.209}\\
\bottomrule
\end{tabular}
\end{table}

The results in Table~\ref{t:mttres} show nearly equal tracking accuracy between the non-adaptive and adaptive multi-fidelity cases in each scenario, while the low-fidelity cases see much higher average tracking errors.
In the combined case, the computational cost of track prediction is roughly equal, on average, between the three propagation methods, while the adaptive method yields almost half the update time of the non-adaptive method.
In the \ac{llo} case, the prediction time for the adaptive multi-fidelity method is somewhat higher than that of the non-adaptive method.
This is because the nominal trajectory is below \SI{2000}{\km}, so the adaptive method is mostly using the reduced time step size of $\frac{1}{4}\Delta t$ for low-fidelity prediction to ensure that the minimum altitude is calculated accurately.
However, the slight difference in prediction times is insignificant compared to the difference in update time between the two multi-fidelity methods in the \ac{llo} scenario.

\begin{figure}[htbp]
\centering
\includegraphics[scale = 0.67]{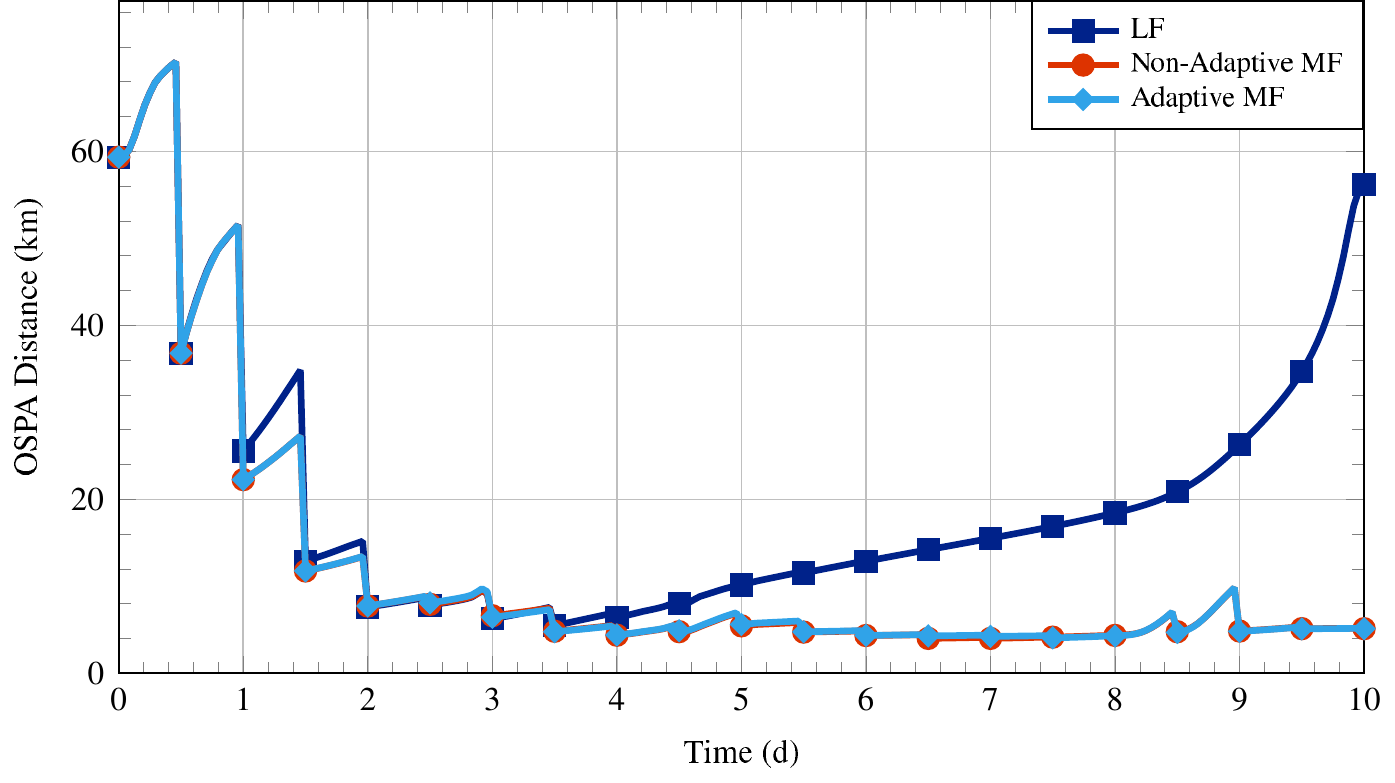}
\caption{Tracking error over time for the combined scenario. Markers coincide with measurement updates.}\label{f:ospacombined}
\end{figure}

Figure~\ref{f:ospacombined} shows the \ac{glmbf}'s tracking error using each method in the combined scenario.
As the figure shows, all three methods initially perform well, but the low-fidelity method begins to diverge around four days into the simulation.
This is primarily due to difficulty in maintaining custody of the objects in the \ac{nrho} and flyby clusters, due to the strong effect of non-spherical gravity at the time of closest approach to the Moon.
On the other hand, both the non-adaptive and adaptive methods have almost the same tracking error over the course of the simulation.
The small spike in error around nine days into the simulation is caused by the \ac{lto} cluster passing close to the Earth, as evidenced by Fig.~\ref{fig:degord_time}.

\begin{figure}[htbp]
\centering
\includegraphics[scale = 0.67]{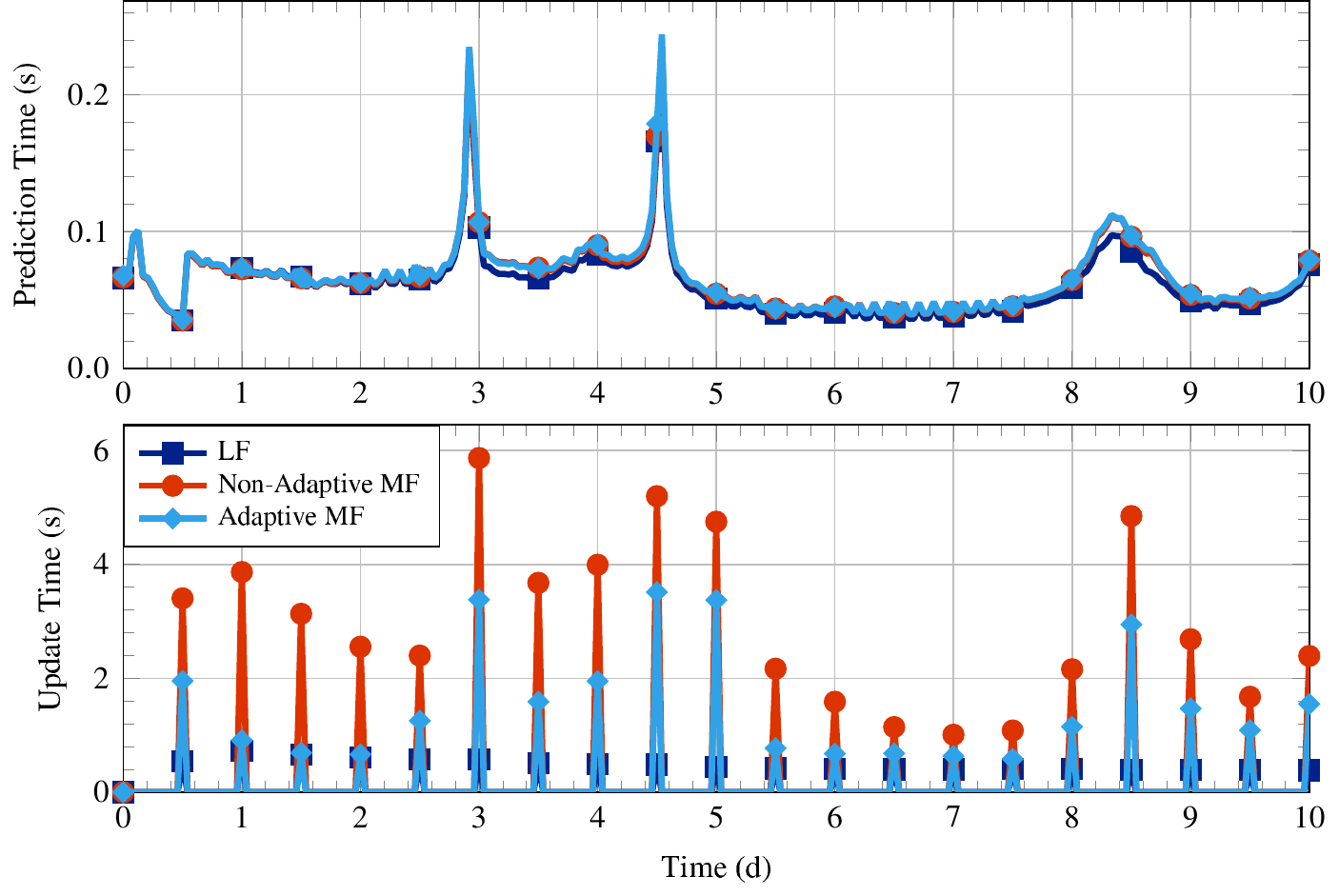}
\caption{Computational cost of each filter step for the combined scenario, broken down by GLMBF algorithm stage.
Markers coincide with measurement updates.}\label{f:timecombined}
\end{figure}

Figure~\ref{f:timecombined} breaks down the runtime cost between track prediction and the rest of the filter's update step at each simulation time step in the combined scenario.
The prediction time plots are nearly identical for all three methods in this scenario, as we would expect given the results in Table~\ref{t:mttres}.
The figure shows that update time is generally zero, with a spike every $12\Delta t$ when measurements are taken.
These spikes are consistently small in the low-fidelity case but vary more with the multi-fidelity methods.
The higher spikes in update time correspond to times when one or more clusters are close to one of the bodies, while the lower spikes correspond to times when the clusters are farther away.
The update time plot for the adaptive method can be seen to approach that of the low-fidelity method in the latter case, which occurs over the first two days and again around seven days into the simulation.
This happens because the recommended degree and order for most tracks falls below the 5$\times$5 threshold during these windows (compare Fig.~\ref{fig:degord_time}), so few, if any, tracks undergo multi-fidelity correction.

\begin{figure}[htbp]
\centering
\includegraphics[scale = 0.67]{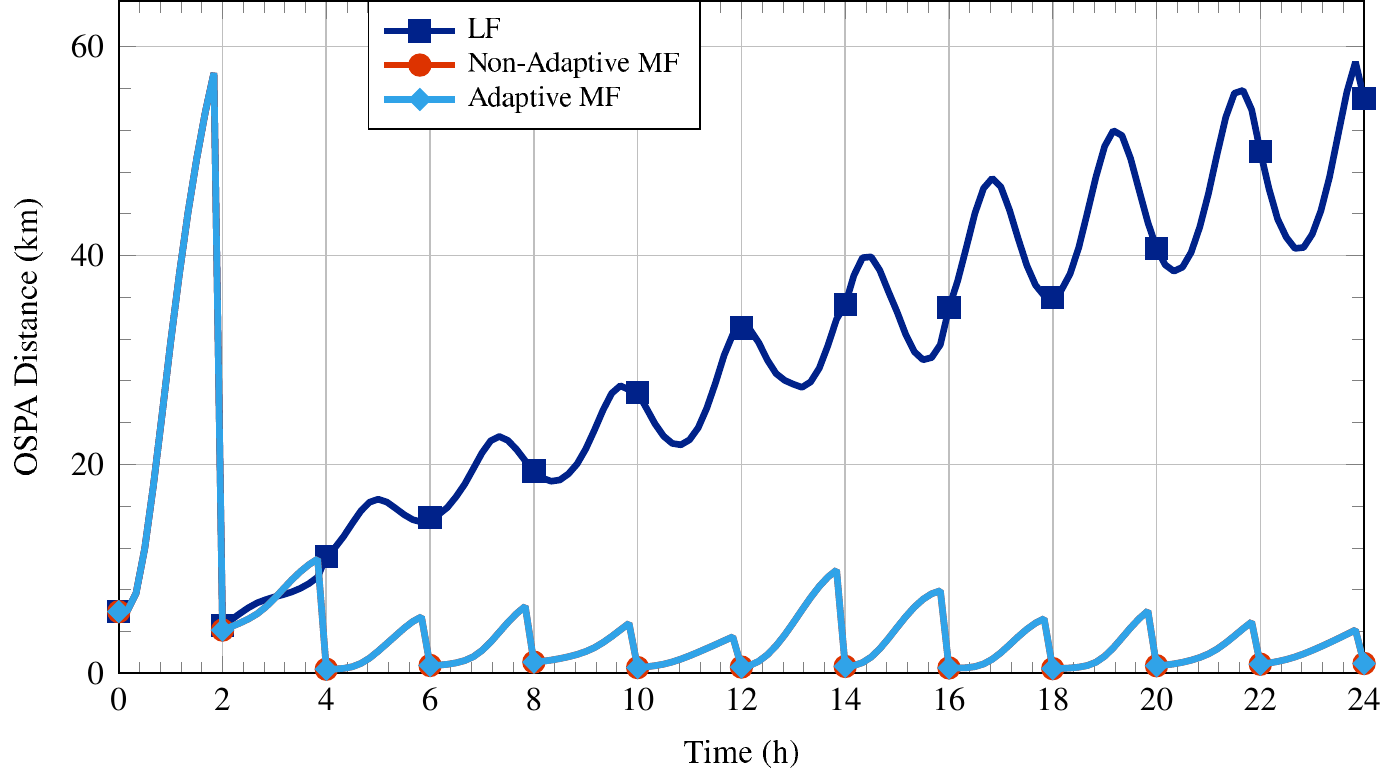}
\caption{Tracking error over time for the LLO scenario. Markers coincide with measurement updates.}\label{f:ospallo}
\end{figure}

\begin{figure}[htbp]
\centering
\includegraphics[scale = 0.67]{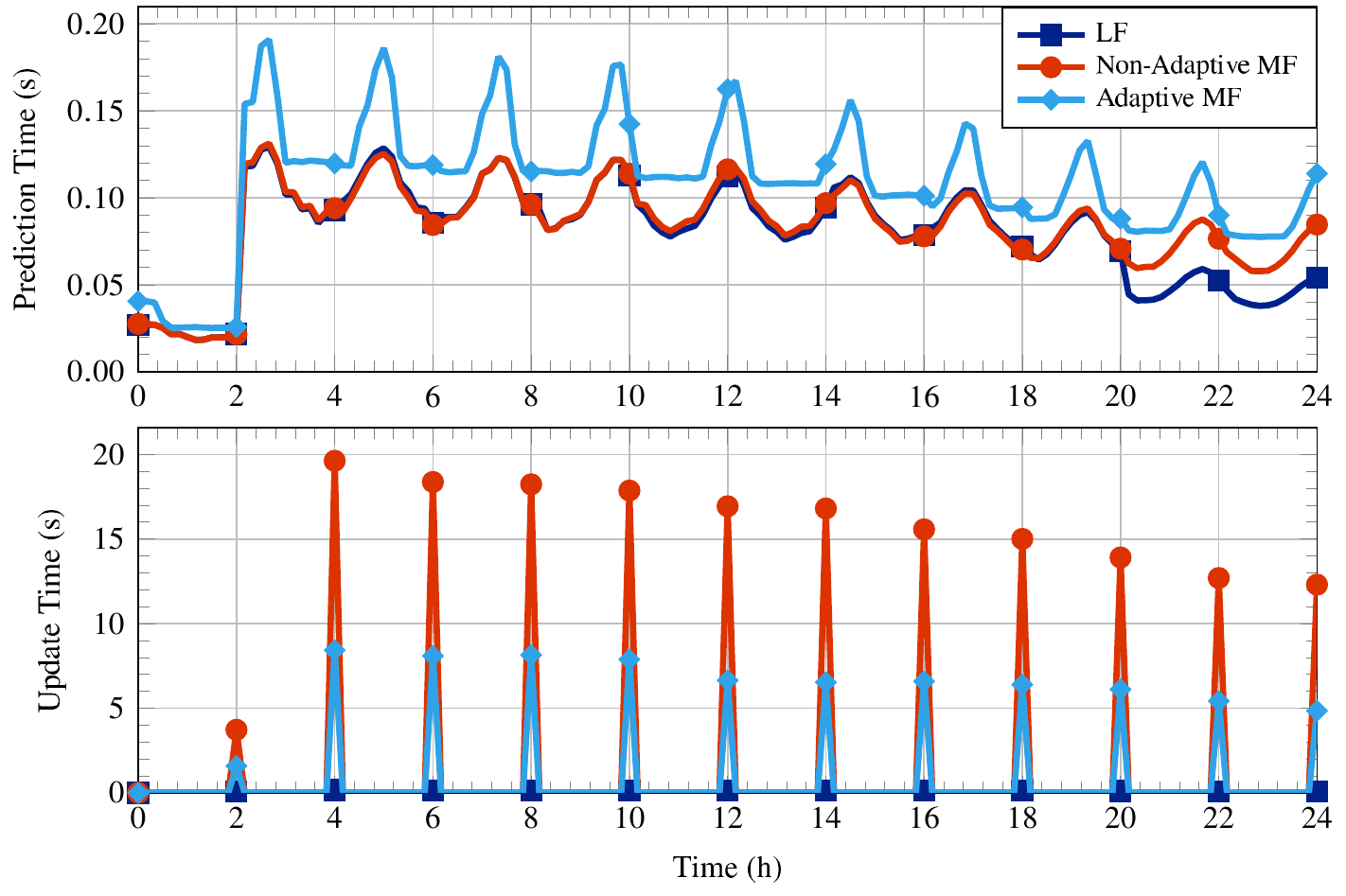}
\caption{Computational cost of each filter step for the LLO scenario, broken down by GLMBF algorithm stage. Markers coincide with measurement updates.}\label{f:timello}
\end{figure}

Figures~\ref{f:ospallo} and \ref{f:timello} show the tracking error and runtime breakdowns for the \ac{llo} scenario.
As in the combined scenario, all three methods initially show similar tracking accuracy, but the low-fidelity method begins to diverge after the first measurement update, two hours into the simulation.
After an initial rise, both the prediction and update time plots are relatively steady, with a slow decrease in computation time over the simulation duration.
This contrasts with Fig.~\ref{fig:degord_time}, where the recommended degree and order slowly increase, due to differing trends in state uncertainty:
without measurements, state uncertainty grows over time, causing the particles parameterizing the state \ac{pdf} to spread out and lowering their minimum altitude.
With measurements, state uncertainty shrinks over time, so the particles parameterizing each track in the \ac{glmbf} tend to converge to the true altitude of the object they represent, raising their minimum altitude.
Notably, in comparison to Fig.~\ref{f:timecombined}, the adaptive method generally yields higher prediction times than the other two methods.
This is driven by the low-fidelity time step size reduction below \SI{2000}{\km}, as noted earlier.
However, once again, the adaptive method yields significantly lower average update times than the non-adaptive method.
This result shows that computational cost of determining the correct degree and order for multi-fidelity correction is less than the cost of assuming a constant degree and order that is known to be sufficient.

\section{Conclusion} \label{sec:conclusion}

This paper developed an adaptive method for multi-fidelity uncertainty propagation in cislunar space by posing the issue as an optimization problem to minimize runtime subject to an upper bound on the expected acceleration error. The solution was obtained by varying the perturbations included in the high-fidelity model as a function of position via a precomputed library of acceleration errors for differing gravity expansion degrees and orders. This permits rapid and accurate uncertainty propagation of space objects in the cislunar domain. The effectiveness of the method for uncertainty propagation was demonstrated via Monte Carlo analysis. For \ac{aso} tracking, the propagator was integrated in a combined \ac{engmf} and \ac{glmbf} multiple-hypothesis, multi-target filter. In simulated test cases, the adaptive approach yielded a significant reduction in runtime when compared to a non-adaptive approach while maintaining equivalent or superior accuracy. Furthermore, the test scenarios demonstrated the method's applicability to orbits relevant to upcoming cislunar missions and \ac{ssa}, including a \ac{dro}, \ac{nrho}, \ac{lto}, low Lunar flyby, and \ac{llo}. The adaptive method also obviates the need for the user to determine an appropriate gravity expansion degree and order, instead determining these automatically to satisfy a specified accuracy requirement. Future work may entail dynamically adjusting the time intervals at which the high-fidelity model is updated, allowing for more rapid adaptation in regions with faster-varying dynamics.

\section*{Funding Sources}
This material is based on research sponsored by the Air Force Research Laboratory (AFRL) under agreement number FA9453-21-2-0064. The U.S. Government is authorized to reproduce and distribute reprints for Governmental purposes notwithstanding any copyright notation thereon. The views and conclusions contained herein are those of the authors and should not be interpreted as necessarily representing the official policies or endorsements, either expressed or implied, of AFRL and/or the U.S. Government. AFRL Public Release Case Number:~AFRL-2026-3095.


\bibliography{references}

\end{document}